\documentstyle[preprint,aps]{revtex}
\def\mass{\mu}
\def\half{\frac{1}{2}}
\def\pv{\vec{\pi}}
\def\sv{\vec{\sigma}}
\def\rv{\vec{r}}
\def\xv{\vec{x}}
\def\rh{\hat{r}}
\def\lv{\vec{L}}
\def\l2{\lv^2}
\def\J2{$J^2$}

\def\okm{\Omega_{\kappa m}}
\def\okmm{\Omega_{-\kappa m}}

\def\Psijme{\Psi^{jm}_E}

\def\fkme{f^{\kappa m E}}
\def\fkmen{f^{\kappa m E}_N}
\def\fkmes{f^{\kappa m E}_S}
\def\fkmeb{f^{\kappa m E}_B}
\def\gkme{g^{\kappa m E}}

\def\gkmes{g^{\kappa m E}_S}

\def\lcrit{l_{\textstyle crit}}
\def\av{\vec{\alpha}}
\makeatother
\begin{document}
\title{Self-Adjoint Extensions of the Pauli Equation in the Presence of a
Magnetic Monopole}
\author{E. Karat\footnote{E-mail address: karat@mit.edu} \\ and 
\\ M. Schulz\footnote{E-mail address: mschulz@physics.berkeley.edu}}
\medskip
\address{Center for Theoretical Physics,
Laboratory for Nuclear Science
and Department of Physics \\[-1ex]
Massachusetts Institute of Technology, Cambridge, MA ~02139--4307}
\maketitle
\setcounter{page}{1}
\begin{abstract}
\noindent
\baselineskip=16pt
We discuss the Hamiltonian for a nonrelativistic electron with spin in
the presence of an abelian magnetic monopole and note that it is not
self-adjoint in the lowest two angular momentum modes.  We then use
von Neumann's theory of self-adjoint extensions to construct a
self-adjoint operator with the same functional form.  In general, this
operator will have eigenstates in which the lowest two angular
momentum modes mix, thereby removing conservation of angular momentum.
However, consistency with the solutions of the Dirac equation limits the
possibilities such that conservation of angular momentum is restored.
Because the same effect occurs for a spinless particle with a
sufficiently attractive inverse square potential, we also study this
system.  We use this simpler Hamiltonian to compare the eigenfunctions
corresponding to a particular self-adjoint extension with the
eigenfunctions satisfying a boundary condition consistent with
probability conservation.
\end{abstract}
\widetext
\section{Introduction}
   In this article, we first examine the Pauli Equation for an electron in 
the field of a magnetic monopole.  This equation has appeared in the 
literature before, and it is well known that an extension is needed to make 
the Hamiltonian self-adjoint in the $j = 0$ sector\cite{dhoker}.
What seems to have gone 
unnoticed is that, for $eg = \frac{1}{2}$,
the domain to be extended includes the $j = 1$
sector as well.  With the inclusion of this sector, the structure of the 
extensions becomes richer, and the number of parameters required to 
describe them jumps from 1 to 16.  While the parameters may be chosen to be 
consistent with angular momentum conservation, this is not required: a pure
incoming s-wave can come out with a p-wave component, even though the 
functional form of the Hamiltonian is 
spherically symmetric.  However, if we require our states to match the
states of the Dirac equation in the nonrelativistic limit, we will only
have 1 free parameter, and angular momentum will be conserved.
   To better understand the effect of the extension parameters and their 
relation to angular momentum conservation, we also consider a simpler
Hamiltonian of the form
\begin{equation}
\label{simpler hamiltonian}
H = - \frac{1}{2\mass } \nabla^2 - \frac {c}{2 \mass r^2},
\end{equation}
where $c$ is an arbitrary constant.
(Spin is not essential to this discussion and is omitted.)
This Hamiltonian has all the essential features of the monopole Hamiltonian,
so we can use it to investigate how the extension parameters arise.  To
accomplish this, we look at this Hamiltonian
in 3-space minus a sphere of radius $r_0$ around the origin.
We compare the results of imposing a boundary condition consistent with
probability conservation
with the results of creating a self-adjoint
extension.  Finally, we compare the case of a nonzero radius $r_0$ with that of
a zero radius $r_0$.
\section{PAULI EQUATION}
Working in units where the speed of light and Planck's constant are both equal
to one, the Hamiltonian for an electron (with spin) in an electromagnetic field
is 
\begin{equation}
\label{monopole Hamiltonian}
H = \frac{\vec{\pi}^2}{2\mass } - \frac{e}{2\mass } \vec{\sigma}\cdot\vec{B},
\end{equation}
where $\vec{\pi}$ is the kinematic momentum $\vec{p}-e\vec{A}$, $\vec{A}$ is
the vector potential, and $\vec{B}$ is the magnetic field.  For
a point magnetic monopole of strength $g$ sitting at the origin\cite{coleman},
we have
\begin{eqnarray}
\label{vector potential}
\vec{A} & = & \frac{g(1-\cos \theta)}{r \sin \theta} \hat{\phi} \\
\vec{B} & = & \frac{g\hat{r}}{r^2} \\
H & = & \frac{\vec{\pi}^2}{2\mass } -
\frac{eg \vec{\sigma}\cdot\hat{r}}{2\mass r^2},
\end{eqnarray}
where we have chosen a particular gauge to determine $\vec{A}$.
Using
$\vec{L} = \vec{r}\times \vec{\pi} - eg \hat{r}$, the Hamiltonian can
be rewritten as
\begin{equation}
\label{L Hamiltonian}
H = \pv\cdot\rv \frac{1}{2\mass r^2} \rv\cdot\pv +
\frac{\l2-e^2g^2-eg\sv\cdot\rh}{2\mass r^2}.
\end{equation}
Here $\lv$ is the orbital angular momentum satisfying
$[L^i,V^j] = i \epsilon ^{ijk} V^k$
for any spin-independent vector operator.  Note that $[L^i, H] \neq 0$.  The
angular momentum operator that commutes with $H$ is the total angular momentum
$\vec{J} = \vec{L} + \half\vec{\sigma}$.  In appendix I, (\ref{L Hamiltonian})
is
shown to simplify to
\begin{equation}
\label{simplified Hamiltonian}
H = - \frac{1}{2\mass }\frac{1}{r}\frac{\partial^2}{\partial r^2}r +
\frac{K(K+1)}{2\mass r^2}
\end{equation}
with the further definition
\begin{equation}
K = -1 - \vec{r}\times\vec{\pi}\cdot\vec{\sigma}
\end{equation}
\cite{greiner and lifshitz}.
We can find simultaneous eigenstates of $K$, $J^2$, and $J_z$ involving only
angular variables.  Such ``monopole harmonics'' will be denoted by
$\okm (\theta, \phi)$
and satisfy
\begin{eqnarray}
K \okm & = & \kappa \okm \\
J^2 \okm & = & j(j+1) \okm \\
J_z \okm & = & m \okm
\end{eqnarray}
\begin{equation}
\int d\Omega \left| \okm (\theta, \phi) \right| ^2 = 1
\end{equation}
where $j$ takes on the values $0, 1, 2, \dots$ and $\kappa$ is related to $j$
by
\begin{equation}
\kappa = \pm \sqrt{j(j+1)}
\end{equation}
for $eg = \frac{1}{2}$.  (More generally,
$\kappa = \pm\sqrt{(j+\frac{1}{2})^2-e^2g^2}$
with $j = eg - \half, eg + \half, \dots$.  See appendix I.)  Note that for
$j > 0$ (or $j > eg - \half$ in the general case), there are two sets of
$m$-multiplets for each value of $j$,
one corresponding to $\kappa > 0$, and one corresponding to $\kappa < 0$.
In terms of Bessel functions and the monopole harmonics,
the $E > 0$ solutions to the eigenvalue equation $H \psi = E \psi$, are
\begin{eqnarray}
\label{nonsingular solution}
\Psi^{\kappa mE}_N = r^{-\half}J_{\nu_\kappa}(\lambda r) \okm (\theta, \phi) \\
\label{singular solution}
\Psi^{\kappa mE}_S = r^{-\half}Y_{\nu_\kappa}(\lambda r) \okm (\theta, \phi) \\
\lambda = \sqrt{2\mass E}, \hspace{.15in}\nu_\kappa = \left| \kappa + \half
\right| \nonumber
\end{eqnarray}
and the $E < 0$ solutions are
\begin{eqnarray}
\label{bound solution}
\Psi^{\kappa mE}_B = r^{-\half}K_{\nu_\kappa}(\lambda r) \okm (\theta, \phi) \\
\lambda = \sqrt{-2\mass E}, \hspace{.15in}\nu_\kappa = \left| \kappa + \half
\right|. \nonumber
\end{eqnarray}
(There is another set of $E < 0$ solutions which has
$I_{\nu_\kappa}$
instead of 
$K_{\nu_\kappa}$, but they grow exponentially at large distances and need not
be considered.)
At this point we mention that any dependence on the gauge is
contained entirely in the form of the $\okm$.  The radial part of these
solutions and the eigenvalues of the operators are left invariant under
a guage transformation.
The set (\ref{nonsingular solution}) 
(N for Nonsingular) of solutions vanishes at the origin, while the sets
(\ref{singular solution}) (S for Singular) and (\ref{bound solution})
(B for bound) are
singular at the origin.  Although the $\Psi_N$ are not square
integrable over all space, they can be $\delta$-function normalized and are
square integrable over any finite region.  However, the
$\Psi_S$ and $\Psi_B$ are only normalizable ($\delta$-function normalizable
in the case of the $\Psi_S$)
when $\nu_\kappa < 1 \Leftrightarrow \kappa = 0, -\sqrt{2}$ (i.e.
for the $j=0$ singlet and one $j=1$ triplet).  If
$\nu_\kappa \geq 1$, the $\Psi_S$ and $\Psi_B$ are not square integrable over
any region containing the
origin.  (The cutoff $\nu_\kappa = 1$ is equivalent to a coefficient of
$\kappa (\kappa + 1) = \frac{3}{4}$ for the $\frac{1}{r^2}$ term in
(\ref{simplified Hamiltonian}).  This coefficient is a general cutoff for a
$\frac{1}{r^2}$ potential \cite{narnhofer} \cite{reed and simon}
\cite{farhi and gutman}.)
It is tempting to let the Hamiltonian operator (\ref{simplified Hamiltonian})
act on all linear combinations of the normalizable (including
$\delta$-function normalizable) solutions.  However,
this is incompatible with the Hermiticity condition
$(H\phi, \psi) = (\phi, H\psi)$ for all $\phi,\psi$ in the domain of $H$.
Furthermore, we seek a Hamiltonian which is not only Hermitian, but also
self-adjoint, for only then are its eigenfunctions complete.
At this point, we need to become more precise in our usage of the term
operator, from
now on including the domain of an operator as part of the operator's
definition.  Precise definitions of Hermitian and self-adjoint will be
employed, and can be found in appendix II.
To start with, let $H_1$ be the operator given functionally by $H$ in
(\ref{simplified Hamiltonian}), and defined on all functions $\psi$ in the
Hilbert space such that
$H\psi$ is in the Hilbert space.  We expect the Hamiltonian to be Hermitian;
however,
\begin{equation}
\label{hermiticity condition}
(H_1\psi, \phi) = (\psi, H_1\phi) + \lim_{r \rightarrow 0}
\frac{1}{2 \mass} \int r^2 d\Omega \left(\frac{\partial \psi^*}{\partial r}
\phi - \psi^* \frac{\partial \phi}{\partial r} \right)
\end{equation}
from integration by parts, so $H_1$ is not Hermitian.
Next, define $H_2$ to be the operator identical to $H_1$ except that the
domain is further restricted to
\begin{equation}
\label{hermiticity by parts}
\left\{ \phi \in \mbox{dom($H_1$)} \left| \lim_{r \rightarrow 0}
\int r^2 d\Omega \left(\frac{\partial \psi^*}{\partial r}
\phi - \psi^* \frac{\partial \phi}{\partial r} \right. \right) = 0
\hspace{.15in}
\mbox{for all $\psi$ $\in$ dom($H_1$)} \right\}.
\end{equation}
By comparing (\ref{hermiticity condition}) with
(\ref{hermiticity by parts}), $H_2$ is  seen to be Hermitian.
In other words, the domain of $H_2$ is the set of $\phi$ in the domain of $H_1$
such that
\begin{equation}
\label{define h2}
(\psi, H_2 \phi) = (H_1 \psi, \phi)
\end{equation}
for all $\psi$ in the domain of $H_1$, which means that $H_2^\dagger = H_1$.
($H_1$ is the adjoint of $H_2$.)  Since the domain
of $H_1$ is larger than that of $H_2$,
$H_2$ is not self-adjoint;
however, the domain of $H_2$ can be
extended through a method of von Neumann to create an operator that is
self-adjoint.
According to the von Neumann theory of
self-adjoint extensions, we need to look at the number $n_+$ of normalizable 
solutions to the equation $H_1\phi_+ = +i\mass \phi_+$ and the number $n_-$
of
normalizable solutions to the equation $H_1\phi_- = -i\mass \phi_-$.
(Note that the use of $\mass$ is arbitrary and chosen only to provide the
correct units.  Any positive real constant may be used instead.)
We can index these solutions and denote them by $\phi^i_\pm$,
where $i$ ranges from 1 to $n_\pm$.
If
$n_+=n_-\equiv n$, then $H_2$ can be made self-adjoint by introducing the n
vectors
$\phi^i = \phi^i_+ + U^i_j \phi^j_-$ into its domain,
where U is an arbitrary unitary $n \times n$ matrix
(with $n^2$ real parameters).
Thus,
a general element of the domain of the extended Hamiltonian, $H_U$, is of the
form $\sum c_i \phi^i +
\tilde{\phi}$, where $\tilde{\phi}$ is in the domain of $H_2$.
  (The reader can  check that the extended
Hamiltonian satisfies the self adjointness criterion of appendix II.)
For the monopole, the domain of $H_2$ consists of those functions in the
domain of $H_1$ that vanish at the origin at least as fast as $r^{1/2}$.
Then, the normalized
solutions to $H_1\phi_\pm=\pm i\mass \phi_\pm$ are
\begin{eqnarray}
\phi_\pm^{jm}=\sqrt{\frac{8\mass ^2 cos(\nu_\kappa \pi / 2)}
{\pi}}r^{-1/2}K_{\nu_\kappa}
((1\mp i) \mass r) \okm, \\
j = 0, \kappa = 0, m = 0\mbox{ or }j = 1, \kappa = -\sqrt{2}, m = 0, \pm 1.
\nonumber
\end{eqnarray}
There are 4 of each, so von Neumann's Theorem tells us that we need $4^2 = 
16$ parameters to describe each extension.  Given 16 parameters 
in the form of a unitary matrix $U^{jm}_{j'm'}$, $(j,j' = 0,1$;
$m = -j,\dots,j$; $m' = -j',\dots,j')$, the Hamiltonian can be 
made self-adjoint by introducing the 4 vectors
\begin{equation}
\phi^{jm} = \phi_+^{jm} + U^{jm}_{j'm'}\phi_-^{j'm'}
\end{equation}
into its domain.  (On a cautionary note: The
superscripts $jm$ on the $\phi^{jm}$ should be considered merely labels.
The states $\phi^{jm}$ defined above and the $\Psijme$ defined below
are not necessarily eigenstates of angular momentum.)
The $\phi^{jm}$ are not
energy eigenstates; however, for each
$\phi^{jm}$ and for each
positive energy eigenvalue $E$, there exists one energy eigenstate, $\Psijme$,
that differs from $\phi^{jm}$ by an element in the domain of $H_2$.
When $U^{jm}_{j'm'}$ is diagonal, the $\Psijme$ are
simultaneous eigenfunctions of $J^2$, $J_z$, and $H_U$, but in general this is
not the
case:  The $\Psijme$ will be eigenfunctions of $H_U$ only, since the
simultaneous
eigenstates of $H_1$, \J2, and $J_z$ corresponding to $j = 0, 1$ are not in the
domain of $H_U$.  Similarly, not all of the eigenstates of \J2 and $J_z$ will
be
eigenstates of $H_U$.  In other words, a pure angular momentum eigenstate with
$j = 0, 1$ will in time evolve as a superposition of states with mixed angular
momenta.  Thus, angular momentum is not conserved for general $U$.
Now we can construct the energy eigenstates.
Because the $\Psijme$ differ by $\phi^{jm}$ by an element in the domain of
$H_2$, which vanishes at the origin at least as fast as $r^\half$, we only
need to consider the behavior of the solutions (\ref{nonsingular solution}),
(\ref{singular solution}), and (\ref{bound solution}) at the origin.  To get
a particular energy eigenstate, pick
a value of energy $E$ and a particular $\phi^{jm}$.  Then, look at its behavior
and the behavior of the above solutions for small $r$.  Any linear combination
of
solutions whose small $r$ behavior matches the small $r$ behavior of the
particular $\phi^{jm}$ (up to a part that vanishes at least as fast as
$r^\half$) is an eigenvalue of the self-adjoint operator $H_U$.  For positive
values of $E$, there is precisely one energy eigenstate for each $\phi^{jm}$.
However, most negative values of $E$ fail to yield an eigenstate.  When we
try to match the solution to $\phi^{jm}$, we derive the following relation
between the energy $E$ and the diagonal matrix element $U^{jm}_{jm}$:
\begin{equation}
\label{energy of bound state}
E = -\mass \left[ \frac{1 + i^{\nu_\kappa} U^{jm}_{jm}}{i^{\nu_\kappa} +
U^{jm}_{jm}}
\right]^{\frac{1}{\nu_\kappa}},
\end{equation}
where, as before, $\nu_\kappa = \left| \kappa + \half \right|$ and
$\kappa = \pm
\sqrt{j(j+1)}$.
(Specifically, we are only interested in $\kappa = 0$ for $j=0$ and
$\kappa=-\sqrt{2}$ for $j=1$.)
For the bound state to exist, we require that $E$ be real and negative.
The reality condition requires $\left| U^{jm}_{jm} \right|^2 = 1$.  Since $U$
is unitary, this requires the the row and column corresponding to $j$ and $m$
consist entirely of zeros, except for the diagonal element, which must be of
the form $e^{i \theta}$, where $\theta$ is real.  (Note that this also implies
that the bound state must be an angular momentum eigenstate.)
This allows us to simplify
(\ref{energy of bound state}) to
\begin{equation}
E = - \mass \left[ \frac{cos(\pi \nu_\kappa / 2) + cos(\theta)}
{1 + cos(\theta - \pi
\nu_\kappa / 2)} \right]^\frac{1}{\nu_\kappa}.
\end{equation}
So we see that we also have the additional condition on the diagonal element:
\begin{equation}
cos(\theta) \geq - cos (\pi \nu_\kappa / 2).
\end{equation}
There exists one bound state for each
such row; therefore, there can be anywhere from zero to four bound states,
depending on the particular self-adjoint extension chosen.
Now we compare our results with a similar treatment for the Dirac equation
that has already appeared in the literature\cite{goldhaber}\cite{callias}.
We work with the Dirac Hamiltonian
\begin{equation}
H = \av\cdot \pv + \beta \mass
\end{equation}
and in a basis where
\begin{equation}
\av = \left( \begin{array}{cc}
             0 & \sv \\
             \sv & 0
             \end{array} \right) ,
\hspace{.15 in}
\beta = \left( \begin{array}{cc}
               1 & 0 \\
               0 & -1
               \end{array} \right) ,
\end{equation}
and $\sv$ are the Pauli matrices.  The advantage of this basis is that the
Dirac spinor can be separated into upper and lower bispinors, where the
lower bispinor is dropped in the nonrelativistic limit.  When appropriately
separated, the eigenvalue equation $H \psi = E \psi$ becomes
\begin{equation}
\psi = \left( \begin{array}{c} \fkme(r) \okm \\ \gkme(r) \okmm \end{array}
\right)
\end{equation}
\begin{eqnarray}
(\mass - E)\fkme - i(\partial_r + \frac{1-\kappa}{r})\gkme & = & 0 \\
i(\partial_r + \frac{1+\kappa}{r})\fkme + (\mass + E)\gkme & = & 0.
\end{eqnarray}
Solving for $\fkme$,
we get the familiar solutions for $E > \mu$
\begin{eqnarray}
\fkmen = r^{-\half}J_{\nu_\kappa}(\lambda r) \\
\fkmes = r^{-\half}Y_{\nu_\kappa}(\lambda r) \\
\lambda = \sqrt{E^2 - \mass^2}, \hspace{.15in}
\nu_\kappa = \left| \kappa + \half \right| \nonumber
\end{eqnarray}
and for $E < \mu$
\begin{eqnarray}
\fkmeb = r^{-\half}K_{\nu_\kappa}(\lambda r) \\
\lambda = \sqrt{\mass^2 - E^2}, \hspace{.15in}
\nu_\kappa = \left| \kappa + \half \right|. \nonumber
\end{eqnarray}
These are the same solutions as for the Pauli equation, except that $\lambda$
has the different expression above.  However, if we let $E = \mass + E'$ and
identify $E'$ with the nonrelativistic energy, we get
$\lambda = \left| 2 \mass E' + {E'}^2 \right|^{1/2}
\longrightarrow \left| 2 \mass E' \right|^{1/2}$
in the nonrelativistic $E' << \mass$ limit, and we recover the solutions
to the Pauli equation in this limit.
One additional difference between the solutions to the Pauli and Dirac
equations is that we now have an additional function
$\gkme = -i(\partial_r + (1+\kappa)r^{-1})\fkme/(\mass + E)$,
which we require to be square integrable over a finite region containing
the origin.  One may easily check for 
$\kappa = -\sqrt{2}$
that $\gkmes$ is not square integrable over this region,
even though $\fkmes$ is.
Thus, we no longer have the singular solution for
$\kappa = -\sqrt{2}$
as we did in the Pauli case above.  When we look for self-adjoint extensions,
we find that we no longer need an extension for the $j=1$ sector; only the
$j=0$ sector requires an extension.  Thus, angular momentum modes may not mix;
furthermore, we only need one parameter to
specify the extension.  From another point of view, consistency with the
nonrelativistic limit of the Dirac equation requires us to fix 15 of the 16
parameters of the Pauli equation.
\section{GENERAL $\frac{1}{r^2}$ POTENTIAL}
The Pauli equation in the presence of a magnetic monopole is just one example
of
a Hamiltonian where the choice of a self-adjoint extension can lead to
non-conservation of angular momentum.  Looking at the form of
(\ref{simplified Hamiltonian}), we can see that the same essential behavior can
be obtained from a spinless particle with a sufficiently attractive inverse
square potential, as in (\ref{simpler hamiltonian}).
Analysis of this new Hamiltonian is simple and analagous to
the analysis of the magnetic monopole.  Similarly, the arguments in this
section can be  modified to include the monopole
Hamiltonian or any similar Hamiltonian that needs an extension.
With this simpler Hamiltonian, we can investigate how the
extension parameters arise.  To do this, consider a space with a sphere of
radius $r_0$ and centered around the origin removed.  We impose conservation
of probability
at the boundary and seek a relation between the extension parameters and the
boundary conditions.
Now, consider a spinless particle governed by the Hamiltonian
(\ref{simpler hamiltonian}), whose functional form is given by
(using $\lv = -i \rv \times \vec{\nabla}$)
\begin{equation}
\label{general Hamiltonian}
H = -\frac{1}{2\mass }\frac{1}{r}\frac{d^2}{dr^2}r +
\frac{-c + L^2}{2\mass r^2}, r \geq r_0.
\end{equation}
This Hamiltonian has appeared in the literature before. \cite{narnhofer}
The solutions to $H \psi = E \psi$ are
\begin{eqnarray}
\Psi^{lmE}_N = r^{-\frac{1}{2}} J_{\nu_l}(\lambda r) Y^{lm}(\Omega)\\
\Psi^{lmE}_S = r^{-\frac{1}{2}} Y_{\nu_l}(\lambda r) Y^{lm}(\Omega)\\
\Psi^{lmE}_B = r^{-\frac{1}{2}} K_{\nu_l}(\lambda r) Y^{lm}(\Omega)\\
\lambda = \sqrt{2 \mass |E|}\mbox{, }\nu_l = \sqrt{\frac{1}{4} + j(j+1) - c},
\nonumber
\end{eqnarray}
where $\Psi^{lmE}_N$ and $\Psi^{lmE}_S$ are the solutions for $E > 0$ and
$\Psi^{lmE}_B$ is the only solution for $E < 0$.  $\Psi^{lmE}_S$ and
$\Psi^{lmE}_B$ are still singular at the origin;
nevertheless,
because the sphere $r < r_0$ is no longer a part of our space, the singularity
of solutions at $r = 0$ is no longer important:  All of the $E < 0$ solutions
can be normalized, and all of the $E > 0$ solutions can be $\delta$-function
normalized.
Now we wish to select those solutions that are consistent with probability
conservation.
Probability is conserved at the boundary $r = r_0$ if
\begin{equation}
\label{probability conservation condition}
\int d\Omega \left. \left[ \psi^* \frac{\partial}{\partial r} \psi -
\psi \frac{\partial}{\partial r} \psi^* \right] \right|_{r_0} = 0.
\end{equation}
To ensure this, we impose the most general boundary condition at $r = r_0$
that is consistent with (\ref{probability conservation condition}) by
introducing a function $g_{r_0}(\Omega, \Omega')$:
\begin{equation}
\label{general boundary condition}
\left. \frac{\partial}{\partial r}\psi(r, \Omega) \right|_{r_0} =
\int d\Omega g_{r_0}(\Omega, \Omega') \psi(r_0,\Omega')
\end{equation}
with the requirement that $g_{r_0}(\Omega,\Omega')$ is Hermitian, i.e.,
\begin{equation}
\label{g Hermitian}
g^*_{r_0}(\Omega',\Omega) = g_{r_0}(\Omega,\Omega').
\end{equation}
We allow the boundary condition to have a continuous dependence on $r_0$
through the explicit appearance of the subscript $r_0$ in
(\ref{general boundary condition}) and (\ref{g Hermitian}).  If
$g_{r_0}(\Omega,\Omega')$ and $\psi(\xv)$ are expanded in spherical harmonics
as
\begin{eqnarray}
\label{g expanded}
g_{r_0}(\Omega,\Omega') & = &
Y_{lm}(\theta, \phi)
g_{r_0}^{lml'm'}
Y^*_{l'm'}(\theta', \phi')
\\
\label{phi expanded}
\psi(\xv) & = & \psi^{lm}(r) Y_{lm}(\theta, \phi),
\end{eqnarray}
then (\ref{general boundary condition}) and (\ref{g Hermitian}) take on the
matrix form
\begin{eqnarray}
\label{matrix boundary condition}
\left. \frac{d}{dr}\psi^{lm}(r)\right|_{r_0} & = & g_{r_0}^{lml'm'}
\psi^{l'm'}(r_0) \\
{g^{l'm'lm}_{r_0}}^* & = & g^{lml'm'}_{r_0}.
\end{eqnarray}
Now, to construct the eigenfunctions of the Hamiltonian, we simply
take the linear combinations of the above solutions for
a given energy $E$ which are consistent with the boundary condition.
The above procedure is self-contained and distinct from the von Neumann
procedure 
It describes a boundary condition that restricts wavefunctions from the entire 
Hilbert space to a subspace on which probability is conserved at the origin. 
To make contact with the von Neumann procedure, we seek a relation between 
the boundary condition (\ref{matrix boundary condition})
and the unitary matrix that needs to be specified
to apply the von Neumann
theory.  Instead of imposing the boundary condition
(\ref{general boundary condition}) on all solutions to the eigenvalue equation,
we can choose a particular self-adjoint extension $H_U$.
To constuct $H_U$, start with
$H_1$, the operator with the functional form of $H$ in
(\ref{general Hamiltonian})
and domain consisting of functions $\psi$ in the Hilbert space such that
$H\psi$
is in the Hilbert space.
Then, create a new operator $H_2$ as we did in
(\ref{define h2})
and look at the
solutions to the equation $H_1 \phi_\pm = \pm i \mass \phi_\pm$.
This time, we obtain a solution
\begin{equation}
\phi^{lm}_\pm(\xv) = \phi^{lm}_\pm(r) Y_{lm}(\theta, \phi)
\hspace{.2in}\mbox{(no sum)}
\end{equation}
for each $l$,$m$.  Since there are infinitely many $l,m$, 
$n_+$ and $n_-$ are infinite, and we
can create a self-adjoint operator $H_U$ by
extending the domain of $H_2$ to include the infinite
collection of vectors $\{ \phi^{lm}(\xv)\}$, where each vector is of the form
\begin{equation}
\phi^{lm}(\xv) = \phi^{lm}_+(\xv) + U^{lm}_{l'm'} \phi^{l'm'}_-(\xv)
\end{equation}
and where U is an infinite-dimensional unitary matrix.  (Again, the $\phi^{lm}$
are not angular momentum components in the sense of (\ref{g expanded})
and (\ref{phi expanded}).)
We now seek a relationship between $U^{lm}_{l'm'}$ and $g_{r_0}^{lml'm'}$.
Enforcing the hermiticity condition,
$(\phi^{lm}, H_U \psi) = (H_U \phi^{lm}, \psi)$, for all $\psi$ in the
domain of $H_U$,
we have
\begin{eqnarray}
\int_{r \geq r_0} d^3\xv {\phi^{lm}}^*(\xv) H_U \psi(\xv)
= \int_{r \geq r_0} d^3\xv (H_U\phi^{lm}(\xv))^* \psi(\xv) \nonumber \\
\Leftrightarrow {\rm Im} \int d\Omega {\phi^{lm}}^*(\xv)
\left. \frac{\partial \psi(\xv)}{\partial r} \right|_{r_0}
= {\rm Im} \int d\Omega \left. \frac{\partial{\phi^{lm}}^*(\xv)}
{\partial r} \psi(\xv) \right|_{r_0}
\label{Hermitian condition}
\end{eqnarray}
using the explicit functional form of the Hamiltonian,
(\ref{general Hamiltonian}).  Expanding $\psi$ as in (\ref{phi expanded}),
we can
take advantage of the orthogonality of the angular momentum harmonics to
integrate them out, leaving only the radial part of the angular momentum
components:
\begin{eqnarray}
\lefteqn{ \left. \sum_{l'm'} \left( \phi_+^{lm}(r)\delta^{lm}_{l'm'} +
U^{lm}_{l'm'}
\phi^{l'm'}_-(r) \right)^* \frac{d \psi^{l'm'}(r)}{dr} \right|_{r_0}}
\nonumber \\
 & = & \left. \sum_{l'm'} \frac{d}{dr}\left( \phi^{lm}_+(r) \delta^{lm}_{l'm'}
+
U^{lm}_{l'm'} \phi^{l'm'}_-(r) \right)^* \psi^{l'm'}(r) \right|_{r_0}
\hspace{.2in}\mbox{(no sum on $lm$)}.
\label{integrated condition}
\end{eqnarray}
(The ``Im'' of (\ref{Hermitian condition}) has been dropped since the
phase of $\psi$ is arbitrary.)  We define
\begin{equation}
a^{lml'm'}(r) = \left( \phi_+^{lm}(r) \delta^{lm}_{l'm'} + U^{lm}_{l'm'}
\phi_-^{l'm'}(r) \right)^*
\hspace{.2in}\mbox{(no sum).}
\end{equation}
Note that ${a^{lml'm'}}^*(r)$ is the $l'm'$ angular momentum component of
$\phi^{lm}(\xv)$, i.e.,
\begin{equation}
\phi^{lm}(\vec{x}) = {a^{lml'm'}}^*(r) Y_{l'm'}(\Omega).
\end{equation}
Viewing $a$ as a matrix in $lm$ and $l'm'$ which depends on $r$, we can write
(\ref{integrated condition}) as
\begin{equation}
\label{a as matrix}
\left. \frac{d}{dr}\psi^{lm}(r) \right|_{r_0} = \left. \left( a^{-1}(r)
\frac{d}{dr} a(r) \right)^{lml'm'} \psi^{l'm'}(r) \right|_{r_0},
\end{equation}
where the inverse and product inside the parentheses are a matrix inverse (but
not a functional inverse with respect to r) and a
matrix product.  Since this is of the same form as
(\ref{general boundary condition}),
\begin{equation}
\label{link}
g^{lml'm'}_{r_0} = \left. \left( a^{-1}(r) \frac{d}{dr} a(r) \right)^{lml'm'}
\right|_{r_0}.
\end{equation}
Furthermore, if we take $\psi = \phi^{l''m''}$ in (\ref{a as matrix})
(so that $\psi^{lm}(r) = {a^{l''m''lm}}^*(r)$), we have
\begin{eqnarray}
& & \frac{d}{dr} {a^{l''m''lm}}^*(r) = \left( a^{-1}(r) \frac{d}{dr}
a(r) \right)^{lml'm'} {a^{l''m''l'm'}}^*(r) \\
& & \Longleftrightarrow \left( a^{-1}(r) \frac{d}{dr} a(r)
\right)^\dagger = \left( a^{-1}(r) \frac{d}{dr} a(r) \right),
\end{eqnarray}
which shows that $g^{lml'm'}_{r_0}$ is Hermitian.
Equation (\ref{link}) is the desired link between the extension and the
boundary condition.  Now that we have found it, we ask what happens in the
limiting case $r_0 = 0$.  To this effect, first note that,
for a fixed value of $c$, most of the
singular Bessel function solutions cease to be normalizable in this limit.
The same holds for the $\phi^{lm}_\pm(\xv)$:  Since many of the
$\phi^{lm}_\pm(\xv)$
are no longer normalizable, any self-adjoint extension defined in this
section which adds non-normalizable $\phi^{lm}_\pm(\xv)$ to the domain of $H_1$
is no longer valid.
To determine which of the solutions $\phi^{lm}_\pm(\xv)$, $\Psi^{lmE}_N(\xv)$,
and $\Psi^{lmE}_B(\xv)$
are still at least $\delta$-function normalizable, we require, as in the
previous section, that the coefficient of the $\frac{1}{r^2}$ term in the
Hamiltonian be less than $\frac{3}{4}$.  Then, our requirement becomes
$l < \lcrit$, where
\begin{equation}
\lcrit (\lcrit + 1) - c = \frac{3}{4}.
\end{equation}
Those extensions that remain valid have U diagonal for
$l \geq \lcrit$, with entries such that the linear combinations of
$\Psi^{lmE}_N$ and $\Psi^{lmE}_S$ within the domain of $H_U$ are purely
$\Psi^{lmE}_N$ for $l \geq \lcrit$.
At this point, the reader may object that we are including an infinite number
of vectors in these extensions while the von Neumann indices are now finite.
This is not a problem,
since the $\phi^{lm}(\xv)$ with $l \geq \lcrit$ that we are
including already exist within the domain of $H_2$.
Returning to equation (\ref{link}), it follows that the only boundary
conditions admissible in the $r_0 \rightarrow 0$ limit are those for which
$g^{lml'm'}_{r_0}$ is diagonal, except possibly for the block with
$l, l' < \lcrit$ as $r_0 \rightarrow 0$.  One should note that the
relationship between $g_{r_0}$ and $U$ becomes singular as $r_0 \rightarrow 0$
due to the singularities of the $\phi^{lm}_\pm (r)$ in $a_{r_0}^{lml'm'}$.
In general, finite entries in U lead to singular entries in
$g_{r_0 \rightarrow 0}$.  Thus, for $r_0 = 0$, it is more convenient to
describe the domain of the Hamiltonian through U than through a boundary
condition at the origin.  However, if one asks for any physical
description of the choice of extension, the formulation
(\ref{matrix boundary condition}) is
more valuable.  It tells us, for instance how the radial flux leaving
the boundary through the $lm$ channel is related to the probability amplitudes
at the boundary for each channel:
\begin{eqnarray}
{\cal J}_{lm} & \equiv & \left. {\psi^{lm}}^*(r) \frac{d}{dr} \psi^{lm}(r)
\right|_{r_0}
\nonumber \\
& = & \sum_{l'm'} {\psi^{lm}}^*(r_0) g^{lml'm'} \psi^{l'm'}(r_0)
\hspace{.2in}\mbox{(no sum on $lm$)}.
\end{eqnarray}
In summary, we can deal with the singularity at the origin by removing a small
sphere of radius $r_0$ from around the origin.  When we do this, we must
impose a boundary
condition consistent with probability conservation, and we need a periodic
function in two angular variables to describe this.  With certain restrictions
on the boundary condition when $r_0=0$, the angular momentum components of this
function are in direct correspondence with the elements of the unitary matrix
required in the von Neumann theory.  It is in this sense that the boundary
condition is equivalent to a choice of self-adjoint extension, or
alternatively, that the self-adjointness condition is equivalent to probability
conservation at the boundary.
\section*{Appendix I}
We start with a Hamiltonian of the form (\ref{L Hamiltonian})
\begin{equation}
H = (\pv\cdot\rv) \frac{1}{2 \mass r^2} (\rv\cdot\pv) +
\frac{\{ L^2 - e^2 g^2 - eg (\sv\cdot\rh) \} }{2 \mass r^2}.
\end{equation}
Using $-eg \rh \cdot \sv = (\lv - \rv \times \pv) \cdot \sv$, we can rewrite
the contents of the curly br
aces as
\begin{eqnarray}
\{ \} &=& \lv ^2 + \lv \cdot \sv - e^2 g^2 - (\rv \times \pv) \cdot \sv \\
&=& (\lv + \half \sv)^2 - 1 - (\rv \times \pv) \cdot \sv + \frac{1}{4} - e^2
g^2.
\end{eqnarray}
Following the convention of \cite{greiner} and \cite{lifshitz}, we define
$K = -1 - (\rv \times \pv) \cdot \sv$.  For an eigenstate of $J^2$
(where $\vec{J} = \lv + \half \sv$), this then becomes
\begin{equation}
\{ \} = J^2 + K + \frac{1}{4} - e^2 g^2.
\end{equation}
The eigenvalues of $J^2$ will be of the form $j(j+1)$ for
$j = eg - \half, eg + \half, \dots$.
We will show below that the eigenvalues of $K$ are
\begin{equation}
\label{general eigenvalue equation for kappa}
\kappa = \pm \sqrt{(j + \half)^2 - e^2 g^2}.
\end{equation}
Given this relation between the eigenvalues of $K$ and $J^2$, the operator
represented by the terms in curly braces has eigenvalues
$\kappa (\kappa + 1)$, where $\kappa$ is the eigenvalue of $K$.
This demonstrates the equivalence of the terms in (\ref{L Hamiltonian}) and
(\ref{simplified Hamiltonian}).  Next, the
vector potential (\ref{vector potential}) for the magnetic monopole has no
radial component,
so
$\pv \cdot \rv = \vec{p} \cdot \rv$
and
$\rv \cdot \pv = \rv \cdot \vec{p}$,
where $\vec{p}$ is just the mechanical momentum operator given by
$- i \vec{\nabla}$.
Thus, the first term of (\ref{L Hamiltonian}) is just the usual
$- \frac{1}{2 \mass} \frac{1}{r} \frac{\partial ^2}{\partial r^2} r$ term.
To prove (\ref{general eigenvalue equation for kappa}), first write
\begin{equation}
(\rv \times \pv) \times (\rv \times \pv) = (\lv + eg\rv) \times (\lv + eg\rv)
= \lv \times \lv + \frac{eg}{r} (\lv \times \rv + \rv \times \lv).
\end{equation}
Now, $\lv$ satisfies $[L^i, V^j] = i \epsilon^{ijk}V^k$ for any vector
operator that
does not depend on spin coordinates, so the relations $\lv \times \lv = i \lv$
and $\lv \times \rv + \rv \times \lv = 2i \rv$ hold, and
\begin{equation}
(\rv \times \pv) \times (\rv \times \pv) = i (\lv + 2eg\rh).
\end{equation}
If we now square the operator $K$, we see that
\begin{eqnarray}
K^2 &=& 1 + 2 \sv \cdot (\rv \times \pv) + (\sv \cdot (\rv \times \pv))^2 \\
&=& 1 + 2 \sv \cdot (\rv \times \pv) + (\rv \times \pv)^2 + i \sv \cdot
[(\rv \times \pv) \times (\rv \times \pv)]
\end{eqnarray}
using the identity
$(\sv \cdot \vec{A})^2 = A^2 + i \sv \cdot (\vec{A} \times \vec{A})$.
Making the substitution $\rv \times \pv = \lv + eg\rh$
($(\rv \times \pv)^2 = L^2 - e^2 g^2$), this finally becomes
\begin{equation}
\label{K^2 relation}
K^2 = 1 + \sv \cdot \lv + L^2 - e^2 g^2 = J^2 + \frac{1}{4} - e^2 g^2.
\end{equation}
Now, $K$ commutes with $J_z$ since it is a total angular momentum scalar.
Therefore, we can find eigenfunctions $\okm$ of $K$ and $J_z$ with respective
eigenvalues $\kappa$ and $m$.  (Although we do not prove it here, these
eigenfunctions are complete.)  Then, from (\ref{K^2 relation}),
\begin{equation}
\label{eigenvalue containing j squared}
K^2 \okm = (J^2 + \frac{1}{4} - e^2 g^2) \okm,
\end{equation}
which shows that $\okm$ is also an eigenfunction of $J^2$.  Replacing $J^2$
in (\ref{eigenvalue containing j squared})
with its eigenvalue $j(j+1)$, we then arrive at
\begin{equation}
\kappa^2 = j(j+1) + \frac{1}{4} - e^2 g^2 = (j + \half)^2 - e^2 g^2.
\end{equation}
This proves (\ref{general eigenvalue equation for kappa}),
as long as we can show that $\kappa$ takes on both positive
and negative values.  The latter follows from the anticommutation of $K$ with
$(\sv \cdot \rh)$:
\begin{equation}
K (\sv \cdot \rh) + (\sv \cdot \rh) K = 0,
\end{equation}
as the reader may verify.  Then, if $K\okm = \kappa \okm$,
\begin{equation}
\label{proportional}
K (\sv \cdot \rh) \okm = - (\sv \cdot \rh) K \okm = - \kappa (\sv \cdot \rh)
\okm
\end{equation}
(the $m$ value is unchanged through multiplication by $\sv \cdot \rh$).
Furthermore, it can be shown that
\begin{equation}
\label{identity at 0}
(\sv \cdot \rh) \Omega_{0 0} = \Omega_{0 0}.
\end{equation}
So, from (\ref{proportional}) and a phase convention that is consistent with
(\ref{identity at 0}),
\begin{equation}
\Omega_{- \kappa m} = (\sv \cdot \rh) \okm.
\end{equation}
\section*{Appendix II}
The definitions of Hermitian and self-adjoint are\cite{appendix II}:
An operator $H$ is Hermitian if its domain is dense, meaning every state in
the Hilbert space can be arbitrarily well-approximated by states in the domain,
and if
\begin{equation}
(H \psi, \phi) = (\psi, H \phi),
\end{equation}
for every $\phi$ and $\psi$ in the domain of $H$.
The adjoint of a densely defined operator $H$, denoted $H^\dagger$, is
defined for any $\psi$ such that there is an $\eta$ for which
\begin{equation}
(\psi, H \phi) = (\eta, \phi),
\end{equation}
for all $\phi$ in the domain of $H$.  In this event $H^\dagger \psi = \eta$.
$H$ is self-adjoint if $H=H^\dagger$.  A crucial part of this definition is
that
the domains of $H$ and $H^\dagger$ be equal.  For a Hermitian $H$, the domain
of $H^\dagger$ is always at least as big as the domain of $H$, and
$H^\dagger \phi = H \phi$ if $\phi$ is in the domain of $H$.
\section*{Acknowledgements}
We thank E. Farhi for discussions about our work. \\
This work is supported in part by funds provided by the U.S. Department of
Energy (D.O.E.) under cooperative research agreement \#DF-FC02-94ER40818.
This material is based upon work supported under a National Science Foundation
Graduate Research Fellowship and the Undergraduate Research Opportunities
Program at MIT.


\begin{thebibliography}{99}
\bigskip
\frenchspacing
\baselineskip=15pt
\advance\parskip by -2pt
\bibitem{dhoker}
E. D'Hoker and L. Vinet, {\it Phys. Lett.} \textbf {137B} (1984) 72
\bibitem{coleman}
For a review, see
S. Coleman in {\it Magnetic Monopole -- 50 years later\/}, from International
School of Subnuclear Physics, 19th, Erice, Italy (1981) 21 (HUTP-82/A032).
\bibitem{greiner and lifshitz}
It is standard to define an operator of this type when solving the Dirac or
Pauli equation for a spherically symmetric system.  For example, see
\cite{greiner} or \cite{lifshitz}
\bibitem{greiner}
W. Greiner (1994) {\it Relativistic Quantum Mechanics: Wave Equations\/}
(Springer-Verlag, New York), 171
\bibitem{lifshitz}
V. Brestetskii, E. Lifshitz, and L. Pitaevskii (1989)
{\it Quantum Electodynamics\/}
(Pergamon, New York), 130
\bibitem{narnhofer}
H. Narnhofer, {\it Acta Physica Austriaca} \textbf {40} (1970), 306
\bibitem{reed and simon}
M. Reed and B. Simon (1975)
{\it Methods in Mathematical Physics, Vol II\/}
(Academic Press, New York), 159-161
\bibitem{farhi and gutman}
E. Farhi and S. Gutmann, {\it Int. J. Mod. Phys.} \textbf {A5} (1990), 3029
\bibitem{goldhaber}
Yoichi Kazama, Chen Ning Yang, and A. S. Goldhaber, {\it Phys. Rev.}
\textbf {D15} (1977), 2287 \\
A. S. Goldhaber, {\it Phys. Rev.} \textbf {D16} (1977), 1815
\bibitem{callias}
C. J. Callias, {\it Phys. Rev.} \textbf {D16} (1977), 3068
\bibitem{appendix II}
These definitions are taken directly from appendix A of ref.
\cite{farhi and gutman}.
\nonfrenchspacing
\end{thebibliography}
\end{document}